\documentclass[USenglish,twocolumn]{article}

\usepackage[utf8]{inputenc}%(only for the pdftex engine)
\usepackage{microtype}
\usepackage{natbib}
\usepackage{graphicx}
\usepackage{url}
\usepackage{amssymb}
\usepackage[T1]{fontenc}
\usepackage{lmodern}
\begin{document}

\title{Asymmetry between galaxies with different spin patterns: A comparison between COSMOS, SDSS, and Pan-STARRS}

\author{Lior Shamir \\ Kansas State University \\ Manhattan, KS 66502 \\ lshamir@mtu.edu; Phone: +1-785-532-4809}

% Authors, for metadata in PDF

% \runningauthor{Shamir}

\date{}

\maketitle

\abstract{
%The spin pattern of a spiral galaxy is a matter of the perspective of the observer, and therefore galaxies with clockwise spin patterns are expected to be identical in their characteristics to galaxies with counterclockwise spin patterns. 
Previous observations of a large number of galaxies show differences between the photometry of spiral galaxies with clockwise spin patterns and spiral galaxies with counterclockwise spin patterns. In this study the mean magnitude of a large number of clockwise galaxies is compared to the mean magnitude of a large number of counterclockwise galaxies. The observed difference between clockwise and counterclockwise spiral galaxies imaged by the space-based COSMOS survey is compared to the differences between clockwise and counterclockwise galaxies imaged by the Earth-based SDSS and Pan-STARRS around the same field. The annotation of clockwise and counterclockwise galaxies is a fully automatic process that does not involve human intervention or machine learning, and in all experiments both clockwise and counterclockwise galaxies are separated from the same fields. The comparison shows that the same asymmetry was identified by all three telescopes, providing strong evidence that the rotation direction of a spiral galaxy is linked to its luminosity as measured from Earth. Analysis of the luminosity difference using a large number of galaxies from different parts of the sky shows that the difference between clockwise and counterclockwise galaxies changes with the direction of observation, and oriented around an axis. % A subset of SDSS galaxies with spectra does not show a statistically significant difference between the absolute magnitude of clockwise and counterclockwise galaxies.

% such that the photometric asymmetry in one hemisphere is inverse to the photometric asymmetry in the opposite hemisphere.
}

% When normalizing the data by the redshift, the photometric asymmetry is eliminated. However, when normalizing the data by the magnitude, statistically significant differences in the redshift remain. 

\section{Introduction}
\label{introduction}

The spin direction pattern of a spiral galaxy depends on the perspective of the observer, and therefore galaxies with clockwise spin patterns are expected to be identical in their other characteristics to galaxies with counterclockwise spin patterns. However, recent Earth-based observations based on large collections of galaxy images have shown evidence of statistically significant difference between the photometry of spiral galaxies that spin clockwise spiral galaxies that spin counterclockwise \citep{shamir2013color,shamir2016asymmetry,shamir2017colour,shamir2017photometric,shamir2017large}. The photometric difference between spiral galaxies with opposite spin directions was first observed by identifying subtle color differences between clockwise and counterclockwise galaxies \citep{shamir2013color}. A stronger statistical significance of the photometric differences was measured by using machine learning \citep{shamir2016asymmetry}. A machine learning system trained on the photometric variables of clockwise and counterclockwise galaxies was able to predict the spin direction of a galaxy in accuracy much greater than the expected 50\% random accuracy. The probability of having such classification accuracy by mere chance is much smaller than $10^{-5}$. The experiment was repeated with manually and automatically annotated sets of spiral galaxies, leading to very consistent results, with very strong statistical significance \citep{shamir2016asymmetry}. The almost perfect agreement between experiments done with manually classified galaxies and experiments that used automatically annotated galaxies \citep{shamir2016asymmetry} substantially reduces the possibility that the asymmetry is driven by human bias or by a computer error. The dataset was too small to identify specific photometric variables that differentiate between galaxies with opposite spin directions, but it provided strong evidence of a link between the photometry of a galaxy and its spin direction \citep{shamir2016asymmetry}.

Experiments with $\sim1.62\cdot10^5$ galaxies \citep{kuminski2016computer,paul2018catalog} showed significant difference (P$<10^{-5}$) between the brightness of galaxies with clockwise spin patterns and the brightness galaxies with counterclockwise spin patterns \citep{shamir2017photometric}. The experiments also showed that the photometric differences changed with the direction of observation \citep{shamir2017photometric}. % galaxies with clockwise spin patterns are brighter than galaxies with counterclockwise spin patterns in one hemisphere, but dimmer than counterclockwise galaxies in the opposite hemisphere \citep{shamir2017photometric}. 
These observations are consistent across SDSS and Pan-STARRS, both showing the same profile of the asymmetry \citep{shamir2017large}. Repeating the same experiment with manually classified galaxies also showed the same results, providing additional evidence that the asymmetry is not likely the result of a computer error \citep{shamir2017large}. The fact that two different telescopes and two different analysis methods provide the same profile of the asymmetry indicates that the asymmetry is not a feature of an anomaly in a specific instrument or photometric pipeline. The most likely axis around which the asymmetry is oriented is ($\alpha=175^o,\delta=50^o$), which is roughly aligned with the galactic pole at ($\alpha\eqsim192^o,\delta\eqsim27^o$). Data for all of these experiments are publicly available \citep{shamir2016asymmetry,shamir2017colour,shamir2017photometric,shamir2017large}. 

% The luminosity differences between galaxies with clockwise spin patterns and galaxies with counterclockwise spin patterns can also be related to previous observations of asymmetry between the number of clockwise galaxies and the number of counterclockwise galaxies \citep{longo2011detection,shamir2012handedness}. Because the number of galaxies that can be observed from Earth is a function of their luminosity, a consistent difference in luminosity between two types of galaxies can lead to a difference in the number of instances of each type of galaxies as observed from Earth. 

Identifying differences between clockwise and counterclockwise galaxies requires large databases of thousands of galaxies, which became available in the post-information era of astronomy. The first attempts to study a possible asymmetry between clockwise and counterclockwise galaxies using large datasets were done by manual annotation of the galaxy images by their spin patterns \citep{iye1991catalog,land2008galaxy,longo2011detection}. An experiment by \cite{iye1991catalog} was based on manual galaxy classification of 8,287 galaxies, 6,525 of them had an identifiable spin pattern. Comparison of the number of galaxies of each spin pattern did not show statistically significant difference. Another attempt to compare the number of galaxies of each spin pattern was done by the Galaxy Zoo citizen science campaign. By using the substantial human labor available through citizen science, the experiment focused on counting the number of galaxies manually classified by their spin pattern, and did not show statistically significant difference between the number of clockwise and counterclockwise galaxies \citep{land2008galaxy}. However, like in \citep{iye1991catalog}, the experiment was limited by a relatively small dataset of galaxies. The manual annotation of clockwise and counterclockwise galaxies also led to substantial bias, as humans tend to misidentify galaxies that spin counterclockwise as elliptical galaxies \citep{hayes2016nature}, leading to a systematic bias in the dataset. Analysis \citep{shamir2017large} of the average brightness of the clockwise and counterclockwise galaxies classified by Galaxy Zoo as ``superclean'' labels (meaning that 95\% or more of the voters agreed on the annotation) showed that the difference was in agreement with the direction and magnitude of the difference observed in \citep{shamir2017photometric}, but due to the small size of the Galaxy Zoo ``superclean'' dataset no statistically significant difference or indifference could be determined \citep{shamir2017large}. \cite{longo2011detection} used four undergraduate students to sort SDSS galaxies by their spin pattern, and showed possible differences between the number of clockwise and counterclockwise galaxies in the dataset, that responds to the direction of observation. Computer annotation of galaxies provided much larger and more consistent datasets, showing differences in the population of clockwise and counterclockwise galaxies in different parts and the sky \citep{shamir2012handedness} and different redshift ranges \citep{shamir2019cosmological}, also showing multipole alignment of the asymmetry \citep{shamir2019large}.

% {\bf Other studies focused on the distribution of clockwise and counterclockwise galaxies \citep{longo2011detection,shamir2012handedness,lee2019mysterious}, showing patterns among galaxies that are for too distant from each other to have any gravitational interaction.} 

While the previous experiments showed that the brightness and color differences between galaxies with opposite spin patterns exist in the entire sky as observed from Earth \citep{shamir2013color,shamir2016asymmetry,shamir2017colour,shamir2017photometric,shamir2017large}, this paper is focused on a certain part of the sky, but imaged by three different telescopes. Two of these telescopes are ground-based, and one is space-based. The results show that all three telescopes show the same asymmetry.

\section{Data}
\label{data}

% \subsection{The Cosmic Evolution Survey (COSMOS) field}

The dataset is galaxies imaged by the Cosmic Evolution Survey (COSMOS) survey of the Hubble Space Telescope (HST), as well as data collected by the Panoramic Survey Telescope and Rapid Response System (Pan-STARRS) and the Sloan Digital Sky Survey (SDSS). The COSMOS field \citep{scoville2007cosmos,koekemoer2007cosmos,capak2007first} is the largest Hubble Space Telescope survey, covering a field of $\sim$2 square degrees, centered at ($\alpha=150.119^o$, $\delta=2.2058^o$). It is a mosaic of 575 neighboring images taken in a total exposure time of $\sim$1000 hours. % The COMSOS field contains $\sim2\cdot10^6$ galaxies.

% \subsection{Analysis of the galaxy images}

Because the experiment is focused on galaxies with clear identifiable morphology, just extended sources with 5$\sigma$ magnitude or brighter were used. Each galaxy image was separated from the F814W band images using {\it Montage} \citep{berriman2004montage}. That provided a dataset of 114,630 galaxies. The galaxy images were then classified by their spin pattern (clockwise and counterclockwise) as was done in \citep{shamir2016asymmetry,shamir2017colour,shamir2017large,shamir2017photometric}. The classification was done by first applying the Ganalyzer algorithm \citep{shamir2011ganalyzer}. Ganalyzer works by converting each galaxy image into its radial intensity plot, and then detecting the peaks of pixel intensities at different radii from the galaxy center. These peaks correspond to the arms of the galaxy. The sign of the linear regression of these peaks at different radial distances from the center is sensitive to the shape of the arm, and therefore determines the spin pattern of the galaxy. Detailed information about Ganalyzer and its application to the identification of spin patterns of spiral galaxies can be found in \citep{shamir2011ganalyzer,hoehn2014characteristics,shamir2016asymmetry}, and the source code is available through the Astrophysics Source Code Library \citep{ganalyzer_ascl}. An important advantage of Ganalyzer is that it is a model-driven algorithm that works in an intuitive manner that is easy to understand, and it is not based on machine learning methods such as deep convolutional neural networks (DCNN). DCNNs are effective methods in image classification, and have been becoming popular for image classification tasks in the past few years. However, DCNNs are based on non-intuitive automatically-generated set of rules that are difficult to understand. Because these networks are trained with manually annotated ``ground truth'' data, they can capture human biases and result in systematically biased classifiers, leading to a consistent error in the annotation when a high number of galaxies are classified. % DCNNs are also trained in a way that is sensitive to other factors such as differences in the sky background that can affect their behavior in different parts of the sky. For example, if more clockwise galaxies in the training set were taken from a certain sky region, more galaxies in that sky region can be classified as clockwise due to the part of the sky in which they were taken.

After the galaxies were classified automatically, a thorough manual process was applied to ensure that all galaxies are classified correctly \citep{shamir2016asymmetry}. The process of manual classification was repeated twice, such that in the second pass the galaxies were mirrored to ensure that the dataset is not affected by a possible human bias. In the end of that long labor-intensive process, a random set of 200 galaxies were selected randomly, and careful examination of the galaxies showed that all classifications were correct. It is therefore very reasonable to assume that the dataset is clean, and that the galaxies are classified correctly.

\begin{figure}[h]
%\figurenum{<text>}
\includegraphics[scale=0.55]{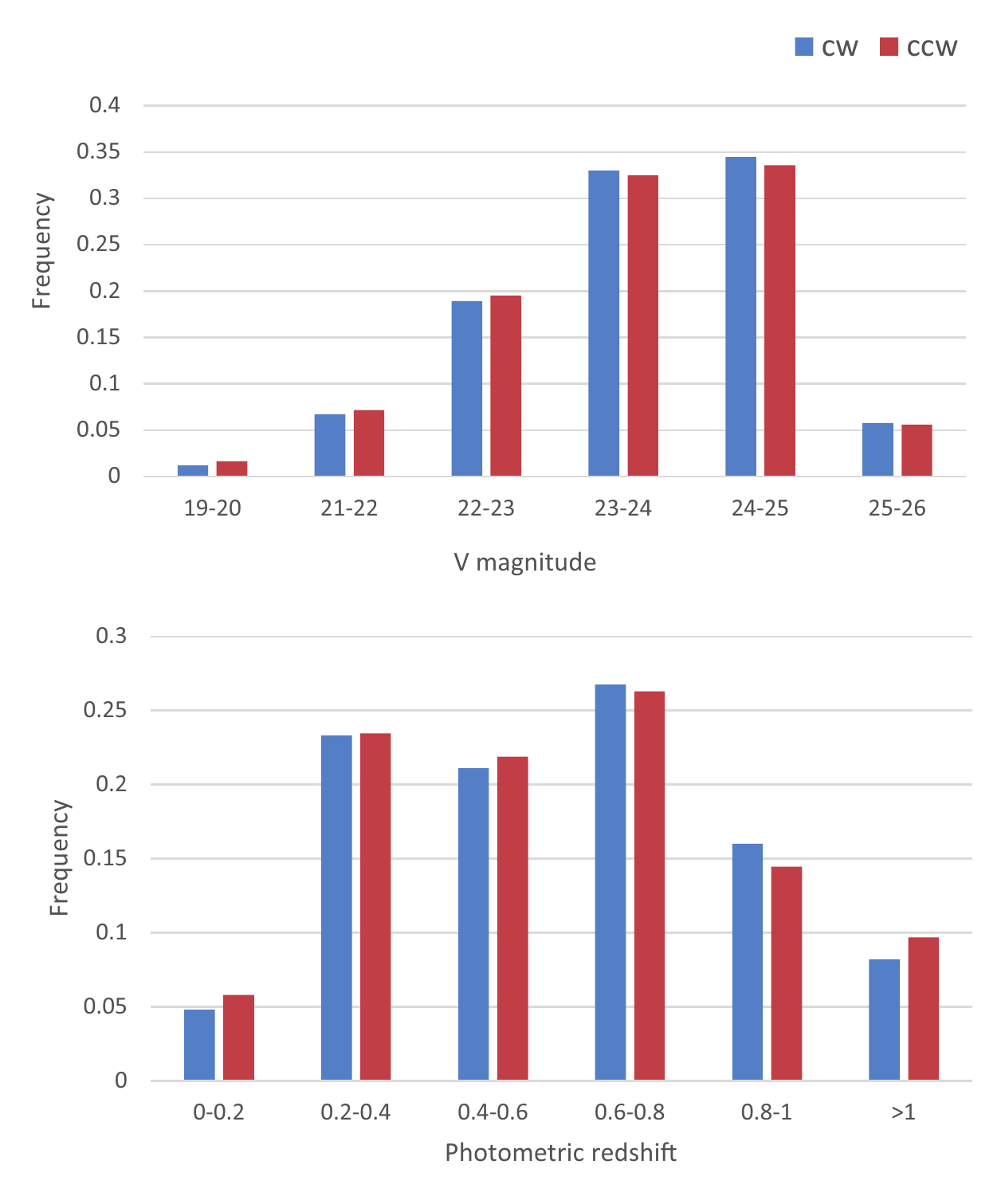}
%\plottwo{<epsfile>}{<epsfile>}
\caption{Histograms of the V magnitude and the photometric redshift of the clockwise and counterclockwise galaxies in the dataset.}
\label{distribution1}
\end{figure}

Since the dataset contained just spiral galaxies with clear spin patterns, the majority of the galaxies were rejected during the process of galaxy classification, as most of them did not have identifiable spin patterns, or were not spiral galaxies. When the process was completed, the dataset contained 5122 galaxies with identifiable spin patterns. These galaxies have mean g magnitude of $\sim$23.1 ($\sigma \simeq 0.97$), V magnitude of $\sim$22.58 ($\sigma \simeq 1.03$), and mean redshift of $\sim$0.6  ($\sigma \simeq 0.28$). These galaxies are clearly brighter than the general COSMOS galaxy population \citep{hasinger2018deimos}, since only galaxies with clear morphology that allows the identification of the spin pattern are used. Fig.~\ref{distribution1} shows the distribution of the V magnitude as well as the photometric redshift of all galaxies in the dataset. The photometric redshift was taken from the photometric redshift catalog of COSMOS \citep{mobasher2007photometric}.

%As the graph shows, some of the galaxies have photometric redshift that is greater than 1. While the photometric redshifts of these objects in the catalog \citep{mobasher2007photometric} are clearly high, all of these objects are spiral galaxies with identifiable spin patterns. Examples of these objects are shown in Table~\ref{highz}. In any case, the photometric z is used here to provide basic information about the galaxy population, and is not used in the analysis.

%\begin{table*}[h]
%\begin{center}
%\caption{COSMOS galaxies with high photometric redshift.}
%\label{highz}
%\begin{tabular}{lccc}
%\hline
%Image & RA ($^o$) & DEC ($^o$)  & PhotoZ  \\
%\hline
%\includegraphics[scale=0.25]{20000363.eps} & 150.672643 & 1.616193 & 1.108 \\
%\includegraphics[scale=0.25]{20003846.eps} & 149.593346 & 1.621579 & 1.284 \\
%\includegraphics[scale=0.25]{20004761.eps} & 150.730916 & 1.656999 & 1.015 \\
%\includegraphics[scale=0.25]{20005731.eps} & 150.628672 & 1.782231 & 1.188 \\
%% \includegraphics[scale=0.25]{20006547.eps} & 150.589341 & 1.672736 & 1.017 \\
%\includegraphics[scale=0.25]{20006580.eps} & 150.585146 & 1.663747 & 1.107 \\
%\hline
%\end{tabular}
%\end{center}
%\end{table*}

% find out what kind of magnitude it is

The dataset of COSMOS galaxies used in this experiment is publicly and freely available at \url{http://people.cs.ksu.edu/~lshamir/data/assym_COSMOS/}. The dataset is separated into two files in the CSV (comma separated values) file format. The first file contains information about clockwise galaxies and the other contains information about counterclockwise galaxies. Each file provides the right ascension and declination coordinates of each galaxy, as well as photometry information.

\section{Results}
\label{results}

The mean magnitudes of the different bands were computed for the 2,607 clockwise COSMOS and for the 2,515 counterclockwise galaxies. Table~\ref{comp1} shows the means and standard errors of the magnitude of the clockwise and counterclockwise galaxies. The table also shows the one-tail P value of the t-test of the difference between the means. The magnitudes are the Subaru AB magnitudes B, V, g, r, i, and z taken from the COSMOS photometry catalog \citep{capak2007first}. A one-tail P value is used here since previous Earth-based observations of SDSS showed difference in the magnitude of clockwise and counterclockwise galaxies in the part of the sky of the COSMOS field \citep{shamir2017photometric,shamir2017large}.

%% T-test is a mature statistical method that tests if the difference between two means is statistically significant. The method determines the probability that the means of two different sets of values are different given the difference between the means, the standard deviation of each mean, and the number of samples in each set \citep{student1908probable}. 

%% Therefore the hypothesis of the experiment is that COSMOS galaxies will show the same asymmetry observed by the ground-based telescopes.

% https://irsa.ipac.caltech.edu/data/COSMOS/gator_docs/cosmos_photom_colDescriptions.html

\begin{table*}[h]
\begin{center}
\caption{Mean, standard error of the mean, and the one-tail statistical significance of the differences between the magnitude of clockwise galaxies and counterclockwise galaxies in COSMOS.}
\label{comp1}
\begin{tabular}{lccc}
Band & mean clockwise & mean counterclockwise & P (t-test)  \\
\hline
B & 23.052$\pm$0.018 & 23$\pm$0.018 & 0.024 \\
V & 22.603$\pm$0.020 & 22.553$\pm$0.02 & 0.042 \\
g & 23.131$\pm$0.019  & 23.077$\pm$0.019 & 0.023 \\
r & 22.266$\pm$0.019 & 22.218$\pm$0.02 & 0.045  \\
i & 21.719$\pm$0.018 &  21.680$\pm$0.018  &  0.065 \\
z & 21.358$\pm$0.017 & 21.323$\pm$0.018 & 0.087 \\
\end{tabular}
\end{center}
\end{table*}

To ensure that no error occurred in the galaxy annotation process, the galaxies were annotated again such that the galaxy images were mirrored, as was done in previous experiments \citep{shamir2012handedness,shamir2016asymmetry,shamir2017colour,shamir2017photometric}. The galaxy images were mirrored using the ImageMagick \citep{still2006definitive} image processing open source tool. Table~\ref{comp_mirrored} shows the results, which are symmetric to the initial experiment with the non-mirrored galaxy images shown in Table~\ref{comp1}.

\begin{table*}[h]
\begin{center}
\caption{Means and standard error of the means of the differences between the magnitude of clockwise galaxies and counterclockwise galaxies in COSMOS when the galaxies were mirrored.}
\label{comp_mirrored}
\begin{tabular}{lccc}
Band & mean clockwise & mean counterclockwise \\
\hline
B & 22.998$\pm$0.018 & 23.052$\pm$0.018 \\
V & 22.551$\pm$0.020 & 22.603$\pm$0.020 \\
g & 23.072$\pm$0.019 & 23.130$\pm$0.019  \\
r & 22.216$\pm$0.020 & 22.263$\pm$0.019 \\
i & 21.660$\pm$0.018  & 21.718$\pm$0.018 \\
z & 21.321$\pm$0.018 & 21.355$\pm$0.017 \\
\end{tabular}
\end{center}
\end{table*}

The mean is sensitive to the distribution of the galaxies, and therefore a small number of bright galaxies in one set can shift the mean and make it different from the mean in the other set. Table~\ref{medians} shows the median values of the magnitude in each of the bands. The table shows that the median magnitude of counterclockwise galaxies is smaller than the median magnitude of clockwise galaxies, in agreement with the differences between the means.

\begin{table*}[h]
\begin{center}
\caption{Median values of the magnitude of clockwise galaxies and counterclockwise galaxies in COSMOS.}
\label{medians}
\begin{tabular}{lccc}
Band & median clockwise & median counterclockwise \\
\hline
B & 23.1440 & 23.1182 \\
V & 22.7483  & 22.7367 \\
g & 23.2456 & 23.2105  \\
r & 22.3602 & 22.3533 \\
i & 21.7858  & 21.7814 \\
z & 21.4497 & 21.4323 \\
\end{tabular}
\end{center}
\end{table*}

Another experiment was done with the same data, such that the values in each set were ordered, and the 5\% smallest values and 5\% largest values of each set were removed. That provided a dataset of 4,610 galaxies, which excludes the 5\% lowest and highest values that can potentially shift the means. Table~\ref{no_tails} shows the means, standard error of the means, and the one-tailed P value of the difference between the means. The results show that removing the 5\% lowest and highest values from both datasets does not change the results significantly, and the difference remains statistically significant.

\begin{table*}[h]
\begin{center}
\caption{Mean, standard error of the mean, and the one-tail P value of the differences between the magnitude of clockwise galaxies and counterclockwise galaxies in COSMOS such that the 5\% highest values and the 5\% lowest values were removed.}
\label{no_tails}
\begin{tabular}{lccc}
Band & mean clockwise & mean counterclockwise & P (t-test)  \\
\hline
B & 23.071$\pm$0.014 & 23.029$\pm$0.015 & 0.027 \\
V & 22.633$\pm$0.017 & 22.586$\pm$0.017 & 0.033 \\
g & 23.156$\pm$0.015  & 23.109$\pm$0.016 & 0.023 \\
r & 22.290$\pm$0.017 & 22.242$\pm$0.017 & 0.031  \\
i & 21.738$\pm$0.015 &  21.697$\pm$0.015  &  0.039 \\
z & 21.382$\pm$0.014 & 21.341$\pm$0.015 & 0.031 \\
\end{tabular}
\end{center}
\end{table*}

The asymmetry detected in COSMOS galaxies was compared to the asymmetry of the galaxies around the same field in SDSS and Pan-STARRS. Since SDSS and Pan-STARRS are less deep than COSMOS, the field was extended to 20$^o$ from the center of the COSMOS field. Therefore, the corners of the field used in SDSS and Pan-STARRS are $(\alpha=130.119^o$, $\delta=22.2058^o)$ and $(\alpha=170.119^o$, $\delta=-17.794^o)$.

The SDSS data were taken from previous work \citep{shamir2017photometric}. The number of SDSS galaxies in that field is 5447 counterclockwise galaxies and 5774 clockwise galaxies. The means and standard errors of the exponential magnitude of the clockwise and counterclockwise galaxies in that field in the different bands are shown in Table~\ref{comp_sdss}. 

%  select count(deVMag_g), avg(deVMag_g), stdev(deVMag_g)/sqrt(count(deVMag_g)) from direction_new_right_phot where ra>=130.119166 and ra<=170.119166 and dec<=22.205833 and dec>=-17.794 and deVMag_g>0 and deVMag_g<25 and deVMag_u>0 and deVMag_u<25 and deVMag_r>0 and deVMag_r<25 and deVMag_i>0 and deVMag_i<25 and deVMag_z>0 and deVMag_z<25    

% \begin{table*}[h]
% \begin{center}
% \caption{Mean DeVaucouleurs magnitude of clockwise and counterclockwise SDSS galaxies in the field centered at the COSMOS field.}
% \label{comp_sdss}
% \begin{tabular}{lccc}
% Band & mean clockwise & mean counterclockwise & P (t-test)  \\
% \hline
% u & 18.2576$\pm$0.014  & 18.2246$\pm$0.015  & 0.050 \\
% g & 17.0752$\pm$0.012  & 17.0359$\pm$0.013 & 0.013 \\ 
% r & 16.4732$\pm$0.012  &  16.4407$\pm$0.012  & 0.028 \\
% i & 16.1513$\pm$0.012 & 16.1206$\pm$0.013 & 0.041 \\
% z & 15.9393$\pm$0.013  &  15.9190$\pm$0.014  & 0.143  \\
% \end{tabular}
% \end{center}
% \end{table*}

% select count(expMag_g), avg(expMag_g), stdev(expMag_g)/sqrt(count(expMag_g)) from direction_new_right_phot where ra>=130.119166 and ra<=170.119166 and dec<=22.205833 and dec>=-17.794 and expMag_g>0 and expMag_g<25 and expMag_u>0 and expMag_u<25 and expMag_r>0 and expMag_r<25 and expMag_i>0 and expMag_i<25 and expMag_z>0 and expMag_z<25

\begin{table*}[h]
\begin{center}
\caption{Mean exponential magnitude of clockwise and counterclockwise SDSS galaxies in the field centered at the COSMOS field.}
\label{comp_sdss}
\begin{tabular}{lccc}
Band & mean clockwise & mean counterclockwise & P (t-test)  \\
\hline
u & 18.8314$\pm$0.013  & 18.8004$\pm$0.012  & 0.043 \\
g & 17.5366$\pm$0.012  & 17.5032$\pm$0.012 & 0.032 \\ 
r & 16.9364$\pm$0.012  &  16.9060$\pm$0.013  & 0.045 \\
i & 16.6114$\pm$0.012 & 16.5820$\pm$0.012 & 0.05 \\
z & 16.4357$\pm$0.013  &  16.4146$\pm$0.014  & 0.142  \\
\end{tabular}
\end{center}
\end{table*}

The table shows statistically significant differences between the exponential magnitude of clockwise galaxies and counterclockwise galaxies, except for the z band. The direction of the difference is aligned with the direction observed in the COSMOS field, in which galaxies with counterclockwise spin patterns were also brighter than galaxies with clockwise spin patterns.

A similar experiment was done with Pan-STARRS galaxies, using clockwise and counterclockwise galaxies that were used in a previous experiment \citep{shamir2017large}. The number of Pan-STARRS galaxies in the same field was 1444 clockwise galaxies and 1438 counterclockwise galaxies, and 21\% of them are also present in the SDSS dataset of the same field. The average Kron magnitudes of the galaxies with clockwise spin patterns and counterclockwise spin patterns are shown in Table~\ref{comp_panstarrs}. Like COSMOS and SDSS, Pan-STARRS also shows statistically significant differences between the brightness of galaxies with clockwise spin patterns and galaxies with counterclockwise spin patterns.

% select into tempz avg(zKronMag), stdev(zKronMag)/sqrt(count(zKronMag)), count(zKronMag) from StackObjectThin,MyDB.combined_ccw_tag where ID=objID and zKronMag>0 and zKronMag<22 and gra>=130.119166 and gra<=170.119166 and gdec<=22.205833 and gdec>=-17.794

\begin{table*}[h]
\begin{center}
\caption{Mean Kron magnitude of clockwise and counterclockwise Pan-STARRS galaxies in the field centered at the COSMOS field.}
\label{comp_panstarrs}
\begin{tabular}{lccc}
Band & mean clockwise & mean counterclockwise & P (t-test)  \\
\hline
g & 16.9774$\pm$0.022 & 16.9133$\pm$0.021 & 0.017 \\
r & 16.3708$\pm$0.021 & 16.3118$\pm$0.020 & 0.021 \\
i & 16.0370$\pm$0.021 &  15.9775$\pm$0.019 & 0.018\\ 
z & 15.8757$\pm$0.021 & 15.8117$\pm$0.019 & 0.012 \\ 
\end{tabular}
\end{center}
\end{table*}

\begin{figure}[h]
%\figurenum{<text>}
\includegraphics[scale=0.55]{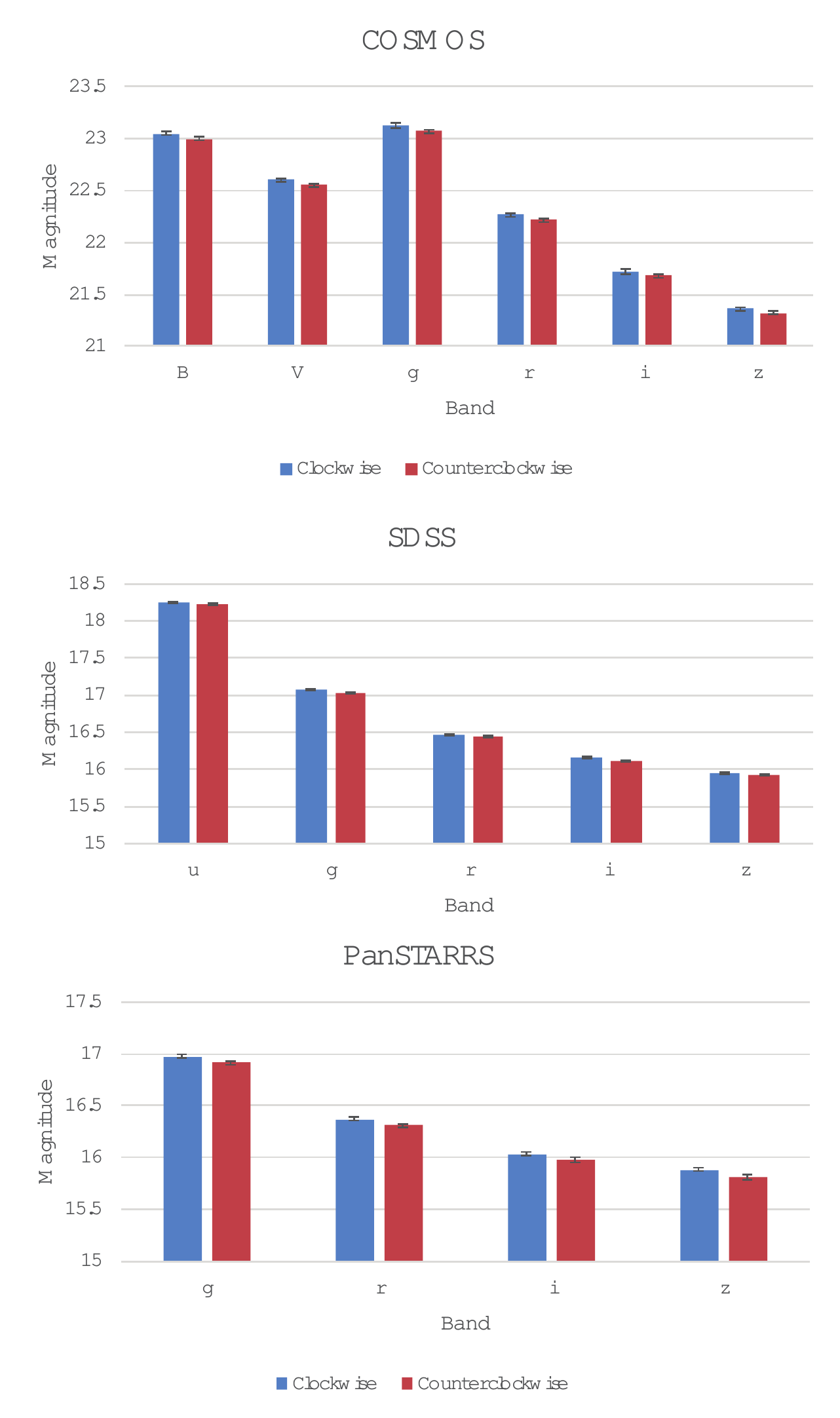}
%\plottwo{<epsfile>}{<epsfile>}
\caption{Mean magnitude and standard error of clockwise and counterclockwise galaxies in the different bands in COSMOS, SDSS and Pan-STARRS.}
\label{comp_fig}
\end{figure}

Figure~\ref{comp_fig} displays the differences between the magnitudes of clockwise and counterclockwise galaxies in each band. The observation that three different telescopes show similar asymmetry reduces the probability that the asymmetry is the result of an anomaly related to a certain telescope or photometric pipeline. Also, the statistical significance of the asymmetry when observed by three telescope is substantially stronger than the asymmetry observed in a single telescope. The mere chance probability that the g magnitude means are different in all three telescopes is $0.023\cdot 0.013\cdot0.017 = \sim5\cdot10^{-5}$. Similarly, the mere chance probabilities that the means are different in all three telescopes in the other bands are $2.6\cdot10^{-5}$, $4.7\cdot10^{-5}$, $1.5\cdot10^{-4}$,  for the r, i, and z bands, respectively.

% Galaxy Zoo

% select count(devMag_u),avg(devMag_u),stdev(devMag_u)/sqrt(count(devMag_u)),avg(devMag_g),stdev(devMag_g)/sqrt(count(devMag_g)),avg(devMag_r),stdev(devMag_r)/sqrt(count(devMag_r)),avg(devMag_i),stdev(devMag_i)/sqrt(count(devMag_i)),avg(devMag_z),stdev(devMag_z)/sqrt(count(devMag_z)) from PhotoObjAll,zooVotes where zooVotes.objid=PhotoObjAll.objid and p_acw>=0.95 and zooVotes.ra>=130 and zooVotes.ra<=170 and zooVotes.dec>-18 and zooVotes.dec<23 and devMag_u>0 and devMag_g>0 and devMag_r>0 and devMag_i>0 and devMag_z>0

% Dev
% acw
%    616 17.201815761529 0.0449883621711537 16.0131653748549 0.0458112902115478 15.4356458419329 0.0451620165130629 15.10243518786 0.0466797142142413 14.8511753531245 0.0452807777023225 
% cw
%  599 17.1660931687522 0.0484348830875441 15.9696013923479 0.045547777542471 15.4192172927729 0.04746976861892 15.1090860685243 0.0519962633275222 14.8788772576639 0.0546082036078009 

%exp
%cw
% 599 17.7187461343552 0.0495449264874954 16.3982596262071 0.048022636310401 15.8386737269432 0.0502421023732802 15.5406787773604 0.0530926227098598 15.4089576899508 0.055839452634029 
%acw
% 616 17.7721792614305 0.0449838023794971 16.4580322132482 0.0474489131515796 15.870018147803 0.0473666716107913 15.5515425886427 0.0474947529887114 15.3967361527604 0.0452519002873834 

\section{Discussion}
\label{conclusion}

This paper shows that galaxy image data from SDSS, Pan-STARRS, and COSMOS (HST) exhibit photometric differences between spiral galaxies the spin clockwise and spiral galaxies that spin counterclockwise. The SDSS and Pan-STARRS galaxies used in this study are positioned around the center of the COSMOS field, to allow comparison with the COSMOS galaxies. When using galaxies from the entire sky footprint of Pan-STARRS and SDSS, the difference between the brightness of galaxies that spin clockwise and galaxies that spin counterclockwise changes with the direction of observation \citep{shamir2017large,shamir2017colour,shamir2017photometric}. 

\begin{figure}[h]
%\figurenum{<text>}
\includegraphics[scale=0.55]{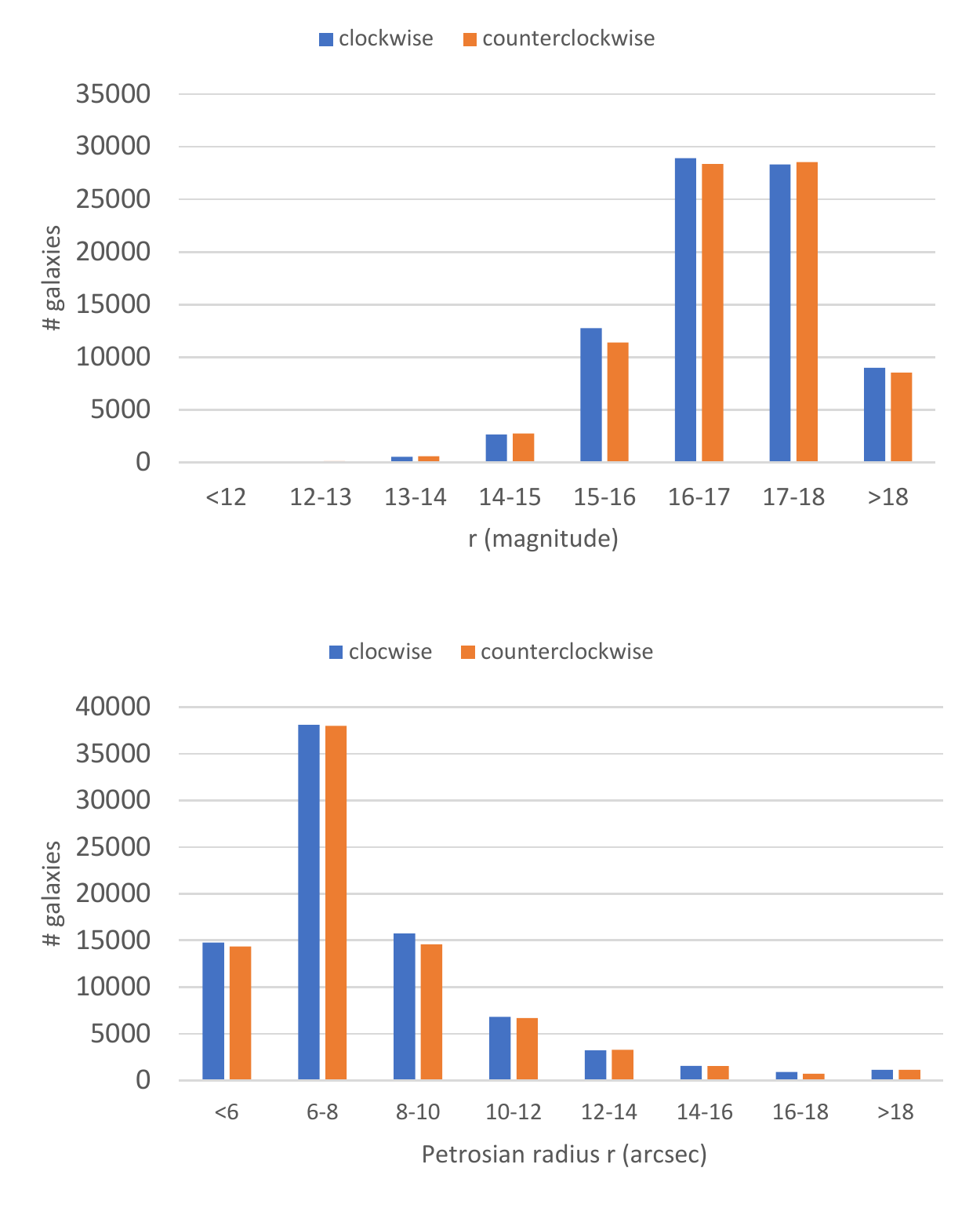}
%\plottwo{<epsfile>}{<epsfile>}
\caption{The distribution of clockwise and counterclockwise galaxies in different r exponential magnitudes and Petrosian radii measured on the r band.}
\label{distribution_SDSS}
\end{figure}

The largest dataset in that experiment was a dataset of 162,516 SDSS galaxies labeled automatically. Figure~\ref{distribution_SDSS} shows the distribution of the radius and exponential r magnitude of the SDSS galaxies. In addition to that dataset, a separate dataset of 40,739 manually labeled spiral galaxies was also used, as well as a dataset of 29,013 Pan-STARRS galaxies labeled automatically. 

Tables~\ref{120_210} and~\ref{300_30} show the mean magnitude of clockwise galaxies and counterclockwise galaxies in the RA range of $(120^o, 210^o)$, as well as the corresponding RA range in the opposite hemisphere $(<30^o ~\vee >300^o)$. The tables also show the t-test P value of the difference between the means of the magnitude of clockwise and counterclockwise galaxies in all three datasets.

\begin{table*}[ht]
\caption{Exponential magnitude mean, standard error, and t-test of the difference between clockwise and counterclockwise galaxies in the $(120^o, 210^o)$ RA range \citep{shamir2017large}.}
\label{120_210}
\begin{center}
{
 \scriptsize 
\begin{tabular}{lcccc}
%\tableline\tableline
Dataset & Band    & Mean & Mean  & t-test  P  \\
            &            & clockwise &  counterclockwise  &     \\   
%\tableline
\hline
SDSS automatically annotated & u & 18.782$\pm$0.006 & 18.757$\pm$0.006 & 0.004 \\
SDSS automatically annotated & g & 17.503$\pm$0.006 & 17.482$\pm$0.006 & 0.016 \\
SDSS automatically annotated & r & 16.913$\pm$0.006 & 16.892$\pm$0.006 & 0.008 \\
SDSS automatically annotated & i & 16.597$\pm$0.006 & 16.578$\pm$0.006 & 0.021 \\
SDSS automatically annotated & z & 16.435$\pm$0.006 & 16.416$\pm$0.006 & 0.033 \\
SDSS manually annotated & u & 18.551$\pm$0.008 & 18.526$\pm$0.008 & 0.033 \\
SDSS manually annotated & g & 17.273$\pm$0.007 & 17.247$\pm$0.008 & 0.022 \\
SDSS manually annotated & r & 16.683$\pm$0.007 & 16.657$\pm$0.008 & 0.013 \\
SDSS manually annotated & i & 16.359$\pm$0.007 & 16.333$\pm$0.008 & 0.014 \\
SDSS manually annotated & z & 16.193$\pm$0.008 & 16.161$\pm$0.008 & 0.003 \\

Pan-STARRS &  g & 17.054$\pm$0.01 & 16.986$\pm$0.01 & $2.8\cdot10^{-5}$ \\
Pan-STARRS &  r & 16.538$\pm$0.01 & 16.471$\pm$0.01 & $1.5\cdot10^{-5}$ \\
Pan-STARRS &  i & 16.236$\pm$0.01 & 16.171$\pm$0.01 & $1.2\cdot10^{-5}$ \\
Pan-STARRS &  z & 16.106$\pm$0.01 & 16.038$\pm$0.01 & $4.6\cdot10^{-6}$ \\
Pan-STARRS &  y & 15.931$\pm$0.01 & 15.897$\pm$0.01 & $7.3\cdot10^{-6}$ \\

%\tableline
\end{tabular}
}
\end{center}
\end{table*}

\begin{table*}[ht]
\caption{Exponential magnitude mean, standard error, and t-test of the difference between clockwise and counterclockwise galaxies in the region of $(<30^o ~\vee >300^o)$.}
\label{300_30}
\begin{center}
{
 \scriptsize
\begin{tabular}{lcccc}
%\tableline\tableline
Dataset & Band    & Mean        & Mean                       & t-test  P  \\
             &              &  clockwise &  counterclockwise   &  (two tails)  \\   
%\tableline
\hline
SDSS automatically annotated &  u & 18.830$\pm$0.007 & 18.883$\pm$0.007 & 3.3$\cdot10^{-8}$ \\
SDSS automatically annotated &  g & 17.508$\pm$0.006 & 17.564$\pm$0.006 & 1.3$\cdot10^{-10}$ \\
SDSS automatically annotated &  r & 16.886$\pm$0.006 & 16.937$\pm$0.006 & 2.1$\cdot10^{-9}$ \\
SDSS automatically annotated &  i & 16.549$\pm$0.006 & 16.601$\pm$0.006 & 2$\cdot10^{-9}$ \\
SDSS automatically annotated &  z & 16.360$\pm$0.006 & 16.415$\pm$0.006 & 2.43$\cdot10^{-9}$ \\
SDSS manually annotated & u & 18.639$\pm$0.02 & 18.648$\pm$0.02 & 0.80 \\
SDSS manually annotated & g & 17.338$\pm$0.02 & 17.347$\pm$0.02 & 0.76 \\
SDSS manually annotated & r & 16.718$\pm$0.02 & 16.738$\pm$0.02 & 0.52 \\
SDSS manually annotated &  i & 16.374$\pm$0.02 & 16.398$\pm$0.02 & 0.44 \\
SDSS manually annotated &  z & 16.193$\pm$0.02 & 16.219$\pm$0.02 & 0.43 \\

Pan-STARRS & g & 17.039$\pm$0.01 & 17.072$\pm$0.01 & 0.085 \\
Pan-STARRS & r & 16.521$\pm$0.01 & 16.550$\pm$0.01 & 0.092 \\
Pan-STARRS & i & 16.195$\pm$0.01 & 16.222$\pm$0.01 & 0.097 \\
Pan-STARRS & z & 16.058$\pm$0.01 & 16.081$\pm$0.01 & 0.187 \\
Pan-STARRS & y & 15.870$\pm$0.01 & 15.896$\pm$0.01 & 0.125 \\

%\tableline
\end{tabular}
}
\end{center}
\end{table*}

\begin{figure}[h]
%\figurenum{<text>}
\includegraphics[scale=0.55]{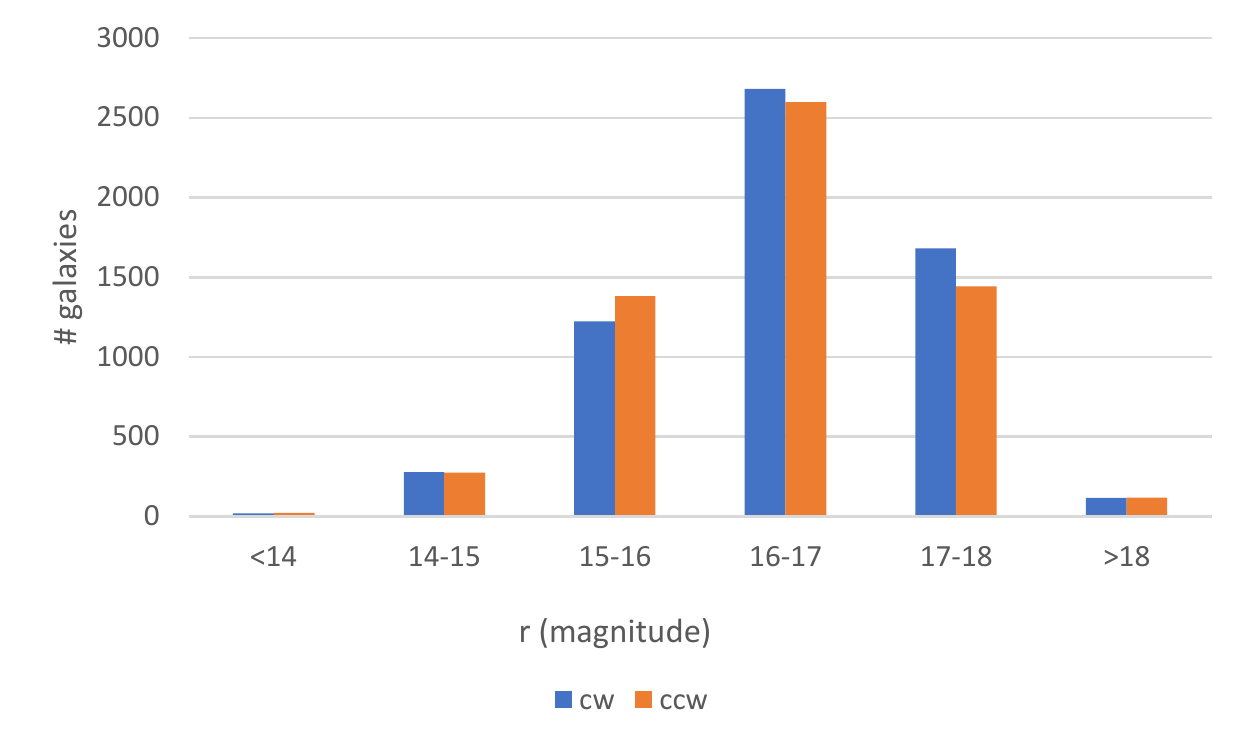}
%\plottwo{<epsfile>}{<epsfile>}
\caption{The distribution of Pan-STARRS clockwise and counterclockwise galaxies in different r exponential magnitudes.}
\label{distribution_PanSTARRS}
\end{figure}

The exponential magnitude model was used, but other magnitude models show similar asymmetry \citep{shamir2017large}. The Student t-test P values are very low in the $(120^o, 210^o)$ RA range due to the much higher number of galaxies. In the RA range $(<30^o ~\vee >300^o)$ there are just 6,597 Pan-STARRS galaxies and 2,889 manually annotated SDSS galaxies, and the P values in that RA ranges are not statistically significant, possibly due to the low number of galaxies. But the automatically annotated SDSS dataset contains 51,053 galaxies, allowing stronger statistical significance. The Pan-STARRS galaxies show a statistically significant difference in the RA range of $(120^o, 210^o)$. The distribution of the clockwise and counterclockwise galaxies in different r exponential magnitudes is shown by Figure~\ref{distribution_PanSTARRS}.

\begin{figure}[h]
\includegraphics[scale=0.65]{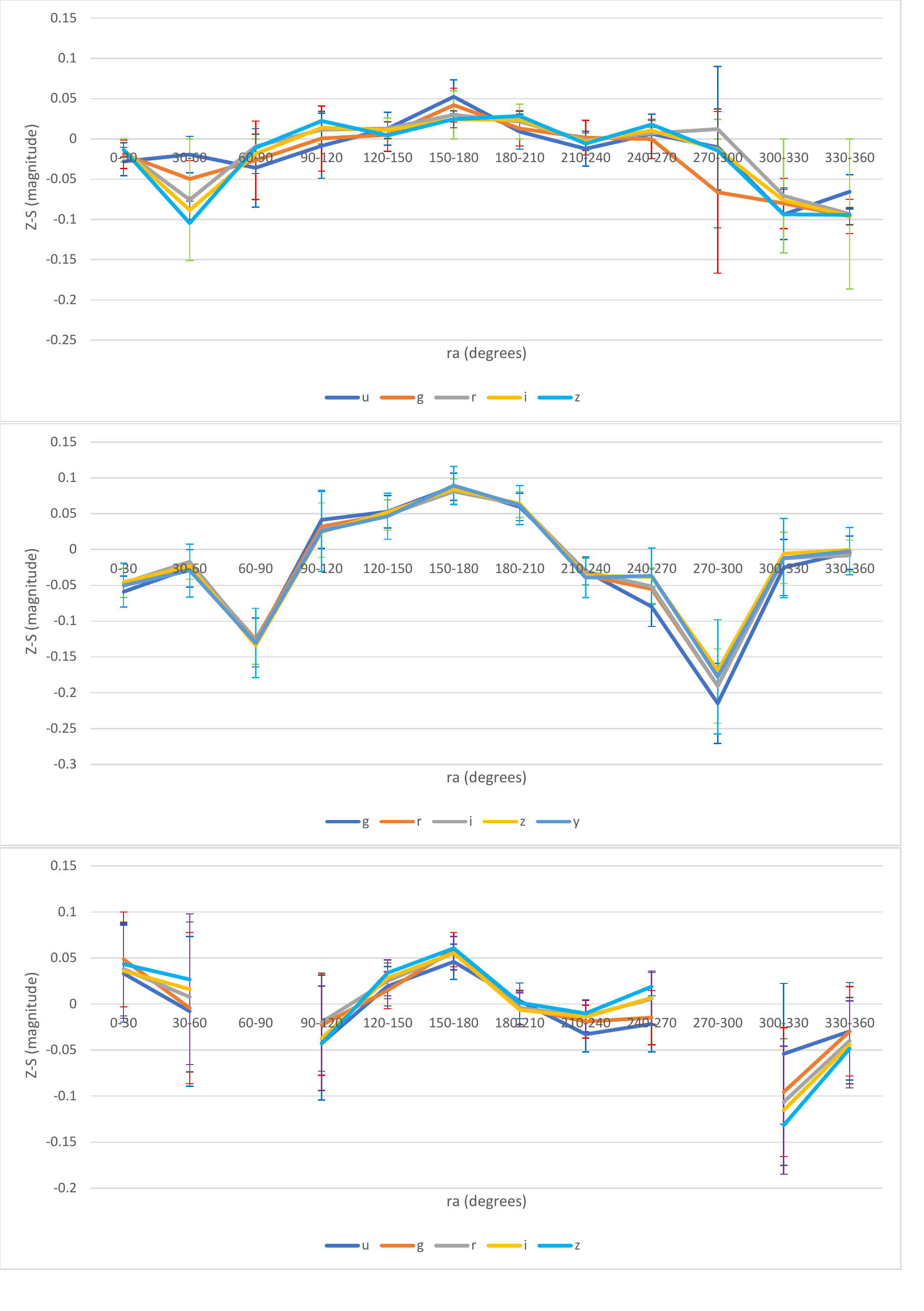}
\caption{Asymmetry between the exponential magnitude of clockwise (Z) and counterclockwise (S) galaxies in different RA ranges using SDSS automatically annotated galaxies (top), SDSS manually annotated galaxies (middle), and Pan-STARRS galaxies \citep{shamir2017large}. The error bars show the standard error of the difference between the mean magnitude of clockwise and counterclockwise galaxies in each band.}
\label{asymmetry_exp}
\end{figure}

Figures~\ref{asymmetry_exp} shows the asymmetry measured using the exponential magnitude in different RA ranges. In the manually annotated SDSS dataset the RA ranges (270$^o$, 300$^o$) and (60$^o$, 90$^o$) were not used because the dataset included galaxies with spectra, which in SDSS are rare in these RA ranges. The figure shows that in all three datasets, the asymmetry changes with the RA, and peaks around the RA range of $(150^o$,180$^o)$.

% The figure shows that in all three datasets, the asymmetry changes with the RA, and peaks around the RA range of $(150^o$,180$^o)$. The figure also shows that the asymmetry in one hemisphere is inverse to the asymmetry in the opposite hemisphere. For instance, in the RA range of $(120^o,210^o)$ counterclockwise galaxies in SDSS are 0.021 g magnitude brighter than clockwise galaxies $(P\simeq0.016)$, while counterclockwise galaxies are 0.056 magnitude dimmer $(P\simeq1.3\cdot10^{-10})$ on the same band in the opposite RA range of $(<30^o \vee>300^o )$. The same is observed in all bands of both SDSS and Pan-STARRS \citep{shamir2017large}.

Since the asymmetry changes with the direction of observation, a flaw in the program that classifies the galaxies \citep{ganalyzer_ascl} is not likely, as such error is expected to exhibit itself in a consistent manner, in the form of similar asymmetry in all parts of the sky. Also, the $\sim4.1\times10^4$ galaxies that were classified manually showed results similar to the results of the automatic classification. These results are also aligned with the experiments described in \citep{shamir2016asymmetry}, where automatically and manually classified galaxies showed nearly identical differences between the magnitude of clockwise and counterclockwise galaxies.

A possible error in the photometric measurements also cannot explain the observed asymmetry. To explain the asymmetry by a photometric measurement error, the error needs to affect clockwise galaxies differently than it effects galaxies with counterclockwise spin patterns. Moreover, the error needs to be systematic, such that the average magnitude of a large number of galaxies is changed by it. Also, since the asymmetry has a very strong and statistically significant correlation with the direction of observation, explaining that observation by a photometric measurement error requires the error to change with the direction of observation. The possibility that all of these three conditions are satisfied is extremely low, and there is no known evidence that even one of these conditions is satisfied.

All three telescopes (SDSS, Pan-STARRS, and HST) show the same asymmetry. Since the imaging conditions might be different for different exposures and different parts of the sky, galaxies imaged at one part of the sky might have different magnitude than galaxies imaged in a different part of the sky or by a different instrument. However, all experiments described here were done such that the comparisons were always done between galaxies within the same part of the sky, imaged by the same instrument, and at the same time. For instance, in the COSMOS field the magnitude of the clockwise galaxies in the $2^2$ degrees field centered at $(\alpha=150.119166^o, \delta=2.205833^o)$ is compared to the magnitude of the counterclockwise galaxies in the $2^2$ degrees field centered at $(\alpha=150.119166^o, \delta=2.205833^o)$. The same is done with SDSS and Pan-STARRS: First the galaxy images are downloaded, and only then they are annotated, separated to clockwise and counterclockwise galaxies, and the magnitudes are compared. Therefore, all comparisons of the magnitudes are done between galaxies in the same field, and no attempt to compare magnitudes in different fields, different instruments, or different exposures is made. Since both the clockwise and the counterclockwise galaxies of each telescope were collected from the exact same field, the difference cannot be the result of cosmic variance \citep{driver2010quantifying}. 

As shown in \citep{shamir2017colour,shamir2017photometric,shamir2017large}, the photometric difference between clockwise and counterclockwise spiral galaxies can be observed in other parts of the sky, but the agreement between Earth-based and space-based observations shows that the asymmetry is not the result of atmospheric effects. The COSMOS field was also imaged in a relatively short time of $\sim$1000 hours, as opposed to SDSS and Pan-STARRS, in which data were acquired over a much longer period of time.

The differences in magnitude between clockwise and counterclockwise spiral galaxies can also be related to previous observations of asymmetry between the number of clockwise galaxies and the number of counterclockwise galaxies \citep{shamir2012handedness}. When galaxies are brighter, more galaxies can be identified and counted. If the apparent magnitude of one type of galaxies is brighter than the other type, that can lead to different frequencies of the two types of galaxies as observed from Earth.

To further examine the link between the possible difference in magnitude and the distribution of clockwise and counterclockwise galaxies, the mean absolute exponential magnitude of clockwise galaxies was compared to the mean absolute exponential magnitude of counterclockwise galaxies. Since COSMOS and Pan-STARRS are not spectroscopic surveys, the comparison was done using the subset of SDSS galaxies used in \citep{shamir2017large} that have spectra. In that dataset, 34,761 galaxies had spectra. Of these galaxies, 17,332 had clockwise spin patterns, and 17,429 had counterclockwise spin patterns. Table~\ref{absolute_mag_all} shows the differences between the absolute exponential magnitude of SDSS clockwise and counterclockwise galaxies with spectra.

\begin{table*}[h]
\begin{center}
\caption{The absolute magnitude of the subset of SDSS galaxies with spectra.}
\label{absolute_mag_all}
\begin{tabular}{lccc}
Band & Mean clockwise & Mean counterclockwise &  P (t-test)  \\
\hline
u &  -18.748$\pm$0.009 & -18.737$\pm$0.008 & 0.373 \\
g &  -20.035$\pm$0.008 & -20.024$\pm$0.008 &  0.344  \\
r & -20.630$\pm$0.009 & -20.618$\pm$0.009 & 0.358 \\
i &  -20.953$\pm$0.009 & -20.939$\pm$0.009 & 0.286 \\
z & -21.127$\pm$0.010 & -21.114$\pm$0.009 & 0.346 \\
\end{tabular}
\end{center}
\end{table*}

The table shows that the differences in the subset of SDSS galaxies with spectra that cover the entire sky imaged by SDSS are not statistically significant. Since the highest asymmetry is observed in the RA range $(120^o,210^o)$, a comparison was made with just galaxies in that part of the sky. The subset of SDSS galaxies with spectra in that sky region included 9,501 galaxies with clockwise spin patterns, and 9,642 galaxies with counterclockwise spin patterns. Table~\ref{absolute_mag_120_210} shows the difference in the absolute exponential magnitude of these galaxies. Like with the subset of SDSS galaxies shown in Table~\ref{absolute_mag_all}, the differences between the absolute magnitude of clockwise and counterclockwise galaxies is not statistically significant. 

\begin{table*}[h]
\begin{center}
\caption{Absolute exponential magnitude of clockwise and counterclockwise SDSS galaxies with spectra in the RA range of $(120^o,210^o)$.}
\label{absolute_mag_120_210}
\begin{tabular}{lccc}
Band & Mean clockwise & Mean counterclockwise &  P (t-test)  \\
\hline
u &  -18.790$\pm$0.012 & -18.794$\pm$0.011 & 0.808 \\
g &   -20.063$\pm$0.011 & -20.067$\pm$0.011 &  0.800 \\
r & -20.651$\pm$0.012 & -20.654$\pm$0.012 & 0.8617 \\
i &  -20.970$\pm$0.013 & -20.971$\pm$0.012 & 0.955 \\
z & -21.143$\pm$0.013 & -21.142$\pm$0.013 &  0.957 \\
\end{tabular}
\end{center}
\end{table*}

The 19,143 SDSS galaxies with spectra in the RA range $(120^o,210^o)$ make a relatively small subset of the entire dataset of SDSS galaxies. To examine the distribution of that subset compared to the entire dataset of SDSS galaxies, the apparent exponential magnitude of clockwise and counterclockwise galaxies in that dataset was compared. Table~\ref{apparent_mag_120_210} shows the differences in the exponential magnitude of clockwise and counterclockwise galaxies of that subset. As the table shows, the differences are also much smaller compared to the differences observed with the entire dataset of SDSS galaxies in the RA range, and the galaxies in that subset are generally brighter than the galaxies in the entire dataset, but due to the much smaller number of galaxies with spectra it is difficult to make a solid comparison between all SDSS galaxies and the subset of SDSS galaxies with spectra.

 % 0.08106$\pm$0.0005 & 0.08110$\pm$0.0005 &
 
\begin{table*}[h]
\begin{center}
\caption{Apparent magnitude of clockwise and counterclockwise SDSS galaxies with spectra in the RA range $(120^o,210^o)$.}
\label{apparent_mag_120_210}
\begin{tabular}{lccc}
Band & Mean clockwise & Mean counterclockwise &  P (t-test)  \\
\hline
u & 18.532$\pm$0.009 & 18.533$\pm$0.009  & 0.93 \\
g & 17.259$\pm$0.009 & 17.259$\pm$0.009  & 1 \\
r & 16.671$\pm$0.008 & 16.673$\pm$0.009  & 0.87 \\
i & 16.351$\pm$0.009 & 16.356$\pm$0.009  & 0.69 \\
z & 16.179$\pm$0.009 & 16.184$\pm$0.009  & 0.69 \\
\end{tabular}
\end{center}
\end{table*}

To examine the same part of the sky in the opposite hemisphere, the subset of SDSS galaxies with spectra was used to compare the absolute exponential magnitude of clockwise and counterclockwise galaxies in the RA range $(<30^o ~\vee >300^o)$, and the results are shown in Table~\ref{absolute_mag_30_300}. The table shows that the clockwise galaxies are somewhat brighter than the counterclockwise galaxies, but the difference is not statistically significant. The dataset is much smaller than the subset of galaxies with spectra in RA range $(120^o,210^o)$, and contains just 1,880 clockwise galaxies and 1,907 counterclockwise galaxies.

\begin{table*}[h]
\begin{center}
\caption{The absolute exponential magnitude of SDSS clockwise and counter clockwise galaxies with spectra in the RA range $(<30^o ~\vee >300^o)$.}
\label{absolute_mag_30_300}
\begin{tabular}{lccc}
Band & Mean clockwise & Mean counterclockwise &  P (t-test)  \\
\hline
u &  -18.625$\pm$0.027 & -18.594$\pm$0.027 & 0.41 \\
g &   -19.951$\pm$0.026 & -19.887$\pm$0.026 &  0.08 \\
r & -20.562$\pm$0.028 & -20.495$\pm$0.027 & 0.08 \\
i &  -20.896$\pm$0.029 & -20.823$\pm$0.029 & 0.07 \\
z &  -21.074$\pm$0.031 & -21.003$\pm$0.030 & 0.1 \\
\end{tabular}
\end{center}
\end{table*}

Using the subset of SDSS galaxies with spectra also allows to observe the distribution of the clockwise and counterclockwise galaxies in different redshift ranges. Tables~\ref{z_120_210} through~\ref{z_210_300} show the number of clockwise and counterclockwise galaxies in different redshift ranges and different parts of the sky. The P values show the binomial probability to have equal or stronger asymmetry between the number of galaxies by chance. The tables do not show statistically significant differences between the distribution of clockwise and counterclockwise galaxies, possibly due to the small size of the sample. The tables show a higher number of counterclockwise galaxies in the RA ranges $(120,^o,210^o)$ and $(<30^o ~\vee >300^o)$, and higher number of clockwise galaxies in the RA ranges $(30^o,120^o)$ and $(210^o,300^o)$. These differences are not statistically significant, and additional work will be required to analyze the distribution of clockwise and counterclockwise galaxies. It should also be noted that the objects with spectra are selected by an algorithm, are are therefore not necessarily a randomly selected sample.

\begin{table}[h]
\begin{center}
\caption{The distribution of clockwise and counterclockwise galaxies in different z ranges in the RA range $(120^o,210^o)$. The P value is the binomial probability of having equal or higher asymmetry between the number of clockwise and counterclockwise galaxies in each redshift range.}
\label{z_120_210}
\begin{tabular}{lccc}
z & cw & ccw & P \\
\hline
0-0.05 & 2566 & 2615   & 0.25 \\  %  -0.009458  5181
0.05-0.1   &  4167 & 4230 & 0.25 \\   %  -0.007503  8397
0.1-0.15  &  2023 & 2036  & 0.42 \\ %    -0.003203  4059
0.15-0.2  &  598 & 614   & 0.33 \\ %   -0.013201  1212
0.2-0.25 &    124 & 121   & 0.39 \\ % 0.39   0.012245  245
0.25- &    23 & 26 & 0.38 \\ %      32
\hline
Total & 9501 & 9642 &  0.15 \\
\end{tabular}
\end{center}
\end{table}

\begin{table}[h]
\begin{center}
\caption{The distribution of clockwise and counterclockwise galaxies in different z ranges in the RA range $(<30^o ~\vee >300^o)$.}
\label{z_30_300}
\begin{tabular}{lccc}
z & cw & ccw & P \\
\hline
0-0.05 &    428 & 462    & 0.13  \\
0.05-0.1 &  919 & 890  & 0.25    \\
0.1-0.15 &   386 & 376   & 0.37  \\
0.15-0.2 &   117 & 136  & 0.13  \\
0.2-0.25 &      20 & 35  & 0.03    \\
0.25-     &  0  & 8  & 0.004 \\
\hline
Total & 1880 & 1907 & 0.33 \\
\end{tabular}
\end{center}
\end{table}

\begin{table}[h]
\begin{center}
\caption{The distribution of clockwise and counterclockwise galaxies in different z ranges in the RA range $(30^o,120^o)$.}
\label{z_30_120}
\begin{tabular}{lccc}
z & cw & ccw & P \\
\hline
0-0.05 &      332 & 314 &  0.25 \\   
0.05-0.1 &      396 & 378 & 0.27 \\   
0.1-0.15 &   208 & 214  & 0.48 \\  
0.15-0.2 &  75 & 59   &  0.09 \\  
0.2-0.25 &    14 & 23  &  0.09 \\  
0.25-    &    6 & 4   &  0.37 \\   
\hline
Total & 1031 & 992 & 0.2 \\
\end{tabular}
\end{center}
\end{table}

\begin{table}[h]
\begin{center}
\caption{The distribution of clockwise and counterclockwise galaxies in different z ranges in the RA range $(210^o,300^o)$.}
\label{z_210_300}
\begin{tabular}{lccc}
z & cw & ccw & P \\
\hline
0-0.05 &     1315 & 1355  & 0.22 \\ %    -0.014981  2670
0.05-0.1 &     2215  & 2212 & 0.49 \\ %       0.000678  4427
0.1-0.15 &    1047 &  963  & 0.03 \\ %       0.041791  2010
0.15-0.2 &  275  & 294  & 0.22 \\ %       -0.033392  569
0.2-0.25 &   59  & 58  & 0.5 \\ %       0.008547  117
0.25-  &   9  &  6  & 0.3 \\ %       0.076923  13
\hline
Total & 4920 & 4888 & 0.37 \\   % 0.37
\end{tabular}
\end{center}
\end{table}

% For instance, as can be seen in Figure 3 in \citep{shamir2012handedness}, the number of clockwise galaxies compared to the number of counterclockwise galaxies peaks in the RA range of 120$^o$-150$^o$, where the luminosity difference between the two types of galaxies is also the highest. However, the number of galaxies is a cruder measurement, as it is based on a boolean value (clockwise or counterclockwise) rather than a measured value.

% An attempt was also made to repeat the experiment with galaxies taken from the Cosmic Assembly Near-infrared Deep Extragalactic Legacy Survey (CANDELS) fields taken by Hubble Space Telescope \citep{grogin2011candels,koekemoer2011candels}: UDS, GOODS-S, GOODS-N, and EGS. However, these fields do not contain a sufficiently large amount of galaxies that have identifiable spin patterns, and therefore did not allow a comprehensive analysis. The numbers of galaxies with annotated spin patterns were 616, 430, 669, and 519 galaxies in the UDS, GOODS-S, GOODS-N, and EGS fields, respectively. These datasets are much smaller than the dataset provided by COSMOS, and do not allow meaningful analysis. 

% \citep{loeb1}

As Figure~\ref{asymmetry_exp} shows, the asymmetry between the brightness of galaxies with opposite directions of rotation changes based on the direction of observation. To identify the axis in which the asymmetry peaks, the two SDSS datasets were combined into a single dataset of galaxies, and the sky was separated into non-overlapping regions with half-width size of 10$^o$ \citep{shamir2017large}. Then, the cosine dependence between the asymmetry in each sky region and the direction of observation was tested for each possible integer combination of ($\alpha$,$\delta$).  % Figure~\ref{correlation_exp} shows the Pearson correlation coefficients of all possible integer ($\alpha$,$\delta$) combinations \citep{shamir2017large}.

% \begin{figure}[h]
% \includegraphics[scale=0.65]{correlation_exp.eps}
% %\plotone{correlation_exp-eps-converted-to.pdf}
% \caption{The Pearson correlation between the exponential magnitude asymmetry and the cosine of the angular distance from all possible integer ($\alpha$,$\delta$) combinations \citep{shamir2017large}.}
% \label{correlation_exp}
% \end{figure}

The highest Pearson correlation coefficient was observed at ($\alpha$=172$^o$,$\delta$=50$^o$), with 1$\sigma$ error range (132$^o$,224$^o$) for the right ascension, and (-26$^o$,74$^o$) for the declination \citep{shamir2017large}. The position of the most likely axis is roughly aligned with the galactic pole at ($\alpha\simeq192^o$, $\delta\simeq27^o$). While it has been proposed that the Cosmic Microwave Background (CMB) is circularly polarized \citep{hu1997cmb,cooray2003cosmic}, the RA of the most likely axis of asymmetry is interestingly close to the RA of the Planck CMB dipole \citep{aghanim2014planck} at 166$^o$. However, there is somewhat substantial difference in the declination (50$^o$ compared to -27$^o$), which is just on the edge of the 1$\sigma$ error range.

% In the majority of the cases, the spin pattern of the galaxy indicate the actual direction in which the galaxy rotates. That is, most galaxies with clockwise spin patterns indeed rotate clockwise. The results of the experiments in this paper show that stars in observed face-on galaxies that rotate in the same direction of the Sun can be measured by an Earth-based system as brighter than stars in face-on galaxies that rotate in the opposite direction. 

The physical nature of galaxy rotation is still not fully understood. Preeminent early astronomers proposed that the galaxy rotation was aligned with Newtonian physics \citep{de1959general,schwarzschild1954mass}, which was in agreement with theory of that time, and played a role in ignoring other observations \citep{slipher1914detection,wolf1914vierteljahresschr,pease1918rotation,babcock1939rotation,mayall1951structure} that showed results that disagreed with the Newtonian physics model of galaxy rotation \citep{rubin2000}. Some early astronomers considered the nature of galaxy rotation as evidence of the existence of nonluminous (``dark'') matter \citep{oort1940some}, but it was only four decades later that the observation became part of ``mainstream'' astronomy \citep{rubin2000one}, accepting the fact that the rotation of galaxies is not driven by Newtonian physics. Since substantial part of the mass of galaxies is not luminous, galaxy rotation is also a tool to measure galaxy mass \citep{sofue2016rotation,sofue2017rotation}. Alternative models to the physics of galaxy rotation such as Conformal Gravity and Modified Newtonian Dynamics (MOND) have also been proposed, and show good agreement with observations \citep{o2017alternative,wojnar2018simple}.

While it is difficult to identify the source of the asymmetry or explain the nature of the observation with current knowledge, the physics of galaxy rotation is not yet understood. One possible explanation could be linked to the large-scale structure of the universe. If the properties of the universe change based on the direction of observation, that would be a violation of the cosmological principle, as such structures that are far larger than any known supercluster violate the isotropy and homogeneity assumptions of the cosmological principle. The clear cosine-dependence between the asymmetry and the direction of observation might be an indication of a rotating universe \citep{godel2000rotating}. The asymmetry between galaxies with opposite spin patterns can be also used as an additional messenger and source of information about the large-scale structure, in addition to other sources of information related to galaxies such as stellar mass growth \citep{alpaslan2016galaxy}, the density distribution, galaxy shape, and the frequency of rare objects \citep{biagetti2019hunt}. The indifference in the absolute magnitude of the galaxies might indicate on differences in the distribution of clockwise and counterclockwise galaxies in different redshift ranges, which can be related to chiral gravitational waves \citep{biagetti2020primordial}. However, the absolute magnitude was measured with a much smaller subset of the galaxies, and no strong conclusions can therefore be made using that smaller dataset. 

However, the source of the asymmetry is not necessarily related to the large-scale structure of the universe, as it can also be related to the internal structure of galaxies or galaxy rotation. One possible explanation for a link between galaxy rotation and the observed asymmetry described in this paper is relativistic beaming. Let $A$ be a face-on spiral galaxy rotating at the same direction as the Milky Way, and $B$ is a face-on spiral galaxy identical to Galaxy $A$, but rotating in the opposite direction of the Milky Way. Both galaxies are in the same hemisphere compared to the Milky Way.

A star moving at rotational velocity $v_r$ relative to the observer will have a Doppler shift of its bolometric flux of $F_o=(1+4\cdot \frac{V_r}{c})$, where $F_0$ is the flux of the star when it is stationary relative to the observer, and $c$ is the speed of light \citep{loeb2003periodic,rybicki2008radiative}. Assuming $\frac{v}{c}$ of the Sun of $\sim$0.001, stars in Galaxy $B$ will have a flux of $F=F_0 (1+4\frac{2\cdot0.001}{c})$, where $F_0$ is the flux of the stars in Galaxy $A$. Therefore, when measuring a large population of galaxies similar to $A$ and $B$, the mean magnitude of the $A$ galaxies will be different from the mean magnitude of the $B$ galaxies.

Let galaxy $C$ be a face-on spiral galaxy rotating in the same direction as galaxy $A$, but located in the opposite hemisphere to $A$ relative to the Milky Way. The spin patterns of a galaxy is normally expected to agree with its actual direction of rotation \citep{iye2019spin}. While galaxies $A$ and $C$ rotate in the same direction, their spin pattern will seem opposite to an Earth-based observer. For instance, if galaxy $A$ would seem a clockwise galaxy to an Earth-based observer, galaxy $C$ would look to an Earth-based observer as a counterclockwise galaxy. Therefore, if clockwise galaxies seem brighter than counterclockwise galaxies in one hemisphere, they are expected to seem dimmer to an Earth-based observer in the opposite hemisphere, since in the opposite hemisphere clockwise galaxies are galaxies that rotate in the opposite direction of clockwise galaxies in the first hemisphere. The asymmetry is expected to peak at the galactic pole, where the rotational velocity relative to observed face-on galaxies peaks.  

If the asymmetry is indeed driven by the relative rotation directions of the observed galaxies compared to the rotation of the Milky Way, the observed differences in the magnitude of each galaxy is relative to the location of the observer. That is, from any galaxy in the universe, face-on spiral galaxies located close to its galactic pole will exhibit different brightness between clockwise and counterclockwise galaxies. That asymmetry is driven by the rotation of these galaxies relative to the galaxy of the observer. Therefore, no cosmological axis that passes through the Earth is needed to explain the observation. The axis where the asymmetry peaks will be located at around the galactic pole of the galaxy of the observer.

The $\frac{F}{F_0}$ calculated above is $\simeq$1.008, and will lead to a maximum magnitude difference of $-2.5\log_{10}1.008 \simeq 0.009$. That difference, however, is smaller than the observed difference in SDSS, Pan-STARRS, and COSMOS. Therefore, while the asymmetry could be related to galaxy rotation, the difference corresponds to a much higher rotational velocity than the observed rotational velocity of galaxies. If the asymmetry is the result of relativistic beaming, it corresponds to a much higher rotational velocity than the actual measured rotational velocity of galaxies.

\section{Acknowledgments}

I would like to thank the two knowledgeable anonymous reviewers for their insightful comments and their help in improving the manuscript. This study was supported in part by NSF grants AST-1903823 and IIS-1546079.

This research is based on observations made with the NASA/ESA Hubble Space Telescope obtained from the Space Telescope Science Institute, which is operated by the Association of Universities for Research in Astronomy, Inc., under NASA contract NAS 5–26555. 

Funding for the Sloan Digital Sky Survey IV has been provided by the Alfred P. Sloan Foundation, the U.S. Department of Energy Office of Science, and the Participating Institutions. SDSS-IV acknowledges support and resources from the Center for High-Performance Computing at the University of Utah. The SDSS web site is www.sdss.org.

SDSS-IV is managed by the Astrophysical Research Consortium for the Participating Institutions of the SDSS Collaboration including the Brazilian Participation Group, the Carnegie Institution for Science, Carnegie Mellon University, the Chilean Participation Group, the French Participation Group, Harvard-Smithsonian Center for Astrophysics, Instituto de Astrof\'isica de Canarias, The Johns Hopkins University, Kavli Institute for the Physics and Mathematics of the Universe (IPMU) / 
University of Tokyo, the Korean Participation Group, Lawrence Berkeley National Laboratory, Leibniz Institut f\"ur Astrophysik Potsdam (AIP), Max-Planck-Institut f\"ur Astronomie (MPIA Heidelberg), Max-Planck-Institut f\"ur Astrophysik (MPA Garching), Max-Planck-Institut f\"ur Extraterrestrische Physik (MPE), National Astronomical Observatories of China, New Mexico State University, New York University, University of Notre Dame, Observat\'ario Nacional / MCTI, The Ohio State University, Pennsylvania State University, Shanghai Astronomical Observatory, United Kingdom Participation Group, Universidad Nacional Aut\'onoma de M\'exico, University of Arizona, University of Colorado Boulder, University of Oxford, University of Portsmouth, University of Utah, University of Virginia, University of Washington, University of Wisconsin, Vanderbilt University, and Yale University.

The Pan-STARRS1 Surveys (PS1) and the PS1 public science archive have been made possible through contributions by the Institute for Astronomy, the University of Hawaii, the Pan-STARRS Project Office, the Max-Planck Society and its participating institutes, the Max Planck Institute for Astronomy, Heidelberg and the Max Planck Institute for Extraterrestrial Physics, Garching, The Johns Hopkins University, Durham University, the University of Edinburgh, the Queen's University Belfast, the Harvard-Smithsonian Center for Astrophysics, the Las Cumbres Observatory Global Telescope Network Incorporated, the National Central University of Taiwan, the Space Telescope Science Institute, the National Aeronautics and Space Administration under Grant No. NNX08AR22G issued through the Planetary Science Division of the NASA Science Mission Directorate, the National Science Foundation Grant No. AST-1238877, the University of Maryland, Eotvos Lorand University (ELTE), the Los Alamos National Laboratory, and the Gordon and Betty Moore Foundation.

\bibliographystyle{apalike}
\bibliography{spin_asymmetry}

\end{document}